\documentclass{elsart}

\usepackage{graphicx}

\usepackage{amssymb}

\journal{Journal of Physics and Chemistry of Solids}
\begin{document}

\begin{frontmatter}



\title{Calculation of valence electron momentum densities using the
  projector augmented-wave method}


\author[HUT]{I. Makkonen\corauthref{cor}},
\corauth[cor]{Corresponding author.}
\ead{ima@fyslab.hut.fi}
\author[Helsinki]{M. Hakala},
\author[HUT]{M. J. Puska}
\address[HUT]{Laboratory of Physics, Helsinki University of Technology, P.O. Box 1100, FIN-02015 HUT, Finland}
\address[Helsinki]{Division of X-Ray Physics, Department of Physical Sciences, P.O. Box 64, FIN-00014 University of Helsinki, Finland}

\begin{abstract}
We present valence electron Compton profiles calculated within the
density-func\-tional theory using the all-electron full-potential
projector augmented-wave method (PAW). Our results for covalent (Si),
metallic (Li, Al) and hydrogen-bonded ((H$_{2}$O)$_{2}$) systems agree well
with experiments and computational results obtained with other
band-structure and basis set schemes. The PAW basis set describes the
high-momen\-tum Fourier components of the valence wave functions
accurately when compared with other basis set schemes and previous
all-electron calculations.
\end{abstract}

\end{frontmatter}


\section{Introduction}

Compton scattering spectroscopy~\cite{Cooper04} is a method
for obtaining direct information on electronic states in
materials. From the scattering cross section one can extract the
Compton profile $J(p_{z})$, which, within the impulse
approximation~\cite{Eisenberger70}, is a one-dimensional projection of the
three-dimensional electron momentum density $\rho(\mathbf{p})$. Compton
scattering experiments are often analyzed with the help of electronic
structure calculations, which are usually based on the framework of
the band theory in the local-density approximation (LDA) of the
density-functional theory~\cite{Martin04} (DFT). A variety of
different methods can be used to represent single-particle states,
charge density and potential in practical calculations. The
approximations made can have a large impact on the quality of the
Compton profiles obtained.

The projector augmented-wave method (PAW) introduced by
P.\ E.\ Bl\"ochl~\cite{Bloechl94} in 1994
is an elegant all-electron method generalizing both the pseudopotential method
and the linearized augmented-plane-wave method (LAPW).
The method includes a well-defined linear transformation from
soft pseudo-valence (PS) wave functions used in the actual calculation to
all-electron (AE) valence wave functions. This is very
practical for the calculation of electron momentum densities,
because the AE wave functions with the rapid oscillations can thus be
easily constructed. One then obtains properly also the high-momentum
Fourier components of the wave functions. The PAW method has already
been applied to many computational problems that require knowledge of
the AE wave functions or the full charge density near the nuclei, for
example, to the calculation of magnetic hyperfine
parameters~\cite{Bloechl00}, electric-field
gradients~\cite{Petrilli98}, and optical properties of
semiconductors~\cite{Adolph01}.

Compton profiles or three-dimensional electron momentum densities have been
calculated in the past with the DFT using, for example, the
full-poten\-tial LAPW (FLAPW)
method~\cite{Blaas95,Kubo95,Kubo97,Baruah99}, localized basis 
sets~\cite{Ghanty00,Hakala04}, pseudowave functions~\cite{Isaacs99},
the orthogonalized plane-wave (OPW) method~\cite{Felsteiner81},
the reconstruction of AE wave functions from pseudowave
functions~\cite{Delaney98}, the
linear muffin-tin orbital method (LMTO)~\cite{Manuel93,Dugdale98,Barbiellini03},
the full-potential LMTO method~\cite{Metz99}, and
the Korringa--Kohn--Rostoker (KKR) band-structure
scheme~\cite{Bansil91,Kaprzyk90,Bansil92}. The Hartree--Fock method has
also been applied~\cite{Ghanty00,Crystal03}, and the $GW$
approximation~\cite{Kubo95,Kubo97,Eguiluz00}, as well as quantum Monte
Carlo (QMC) methods~\cite{Kralik98,Filippi99} which go beyond
the DFT and the LDA. The PAW scheme has
obvious advantages compared with many of these schemes. It is formally
simple and both efficient and flexible in practice. First-row
elements, transition metals and rare-earth elements can be treated. It
uses the full potential and charge density, and AE wave
functions are available. The LMTO and KKR approaches are usually limited to
the muffin-tin form of the potential, and the pseudopotential method
suffers from transferability problems and from the loss of information
on the true wave functions near nuclei. The localized basis sets may have
convergence problems in the interstitial regions of solids, whereas the
basis set of the PAW method is complete and takes into account
the orthogonality of the valence states to the core states
automatically without having to treat also the core electrons
self-consistently. Formally the closest resemblance is to the OPW
method, which also contains a linear transformation between PS and AE
wave functions. Also the reconstruction of AE wave functions from PS
wave functions has some similarities, e.g.\ the existence of soft PS
wave functions and the frozen-core approximation used. However, the PAW
transformation is more effective than the OPW transformation, and the
reconstruction procedure of AE wave functions from PS wave functions is
a more complicated and artificial way to obtain the AE wave functions
than the built-in transformation of the PAW method.

In this work we focus on the PAW method as a computational tool and
demonstrate its applicability to the calculation of the valence electron
momentum densities by calculating Compton profiles for different
systems. According to our knowledge, this is the first time the PAW
method is applied to this problem. Ishibashi~\cite{Ishibashi04} has
recently used the PAW method to calculate momentum densities of annihilating
electron-positron pairs in bulk materials, and Rummukainen \textit{et
  al.}~\cite{Rummukainen04} those in the case of defects is solids. We show
examples of semiconducting, metallic and molecular systems, namely
bulk Si, bulk Li, bulk Al, and the water dimer (H$_{2}$O)$_{2}$, and
compare our results with experiment and other computational
results. We restrict ourselves only to the non-spin-polarized case, but
stress that the PAW method can also be applied to the calculation of
spin-dependent momentum densities, \textit{i.e.}\ magnetic Compton
profiles~\cite{Platzman70,Schulke04,Sakai04}, within the impulse
approximation.


\section{Method of calculation}

\subsection{Projector augmented-wave functions}

The PAW method is based on a linear transformation $\mathcal{T}$
between soft PS valence wave functions $|\tilde{\Psi}\rangle$ and
corresponding AE valence wave functions $|\Psi\rangle$,
\begin{equation}
|\Psi\rangle=\mathcal{T}|\tilde{\Psi}\rangle.
\end{equation}
The transformation differs from unity by local atom-centered
contributions $\hat{\mathcal{T}}_{R}$ such that
\begin{equation}
\mathcal{T}=1+\sum_{R}\hat{\mathcal{T}_{R}}.
\end{equation}
Each local contribution represents the difference between the AE
wave function and the PS wave function and acts within some
augmentation region $\Omega_{R}$ enclosing one atom. The AE
wave function $|\Psi\rangle$ is expanded locally within every
$\Omega_{R}$ into solutions $|\phi_{i}\rangle$ of the Schr\"odinger
equation for the corresponding isolated atom. The index $i$ refers to
the site index $R$, the angular momentum indices $(l,m)$ and an
additional index $k$ referring to the reference energy
$\epsilon_{kl}$. Also the PS wave function $|\tilde{\Psi}\rangle$
is expanded locally using a complete set of soft, fixed PS partial waves
$|\tilde{\phi}_{i}\rangle$ and the same expansion
coefficients as for the AE wave functions. There
is exactly one PS partial wave
$|\tilde{\phi}_{i}\rangle$ for each AE partial wave
$|\phi_{i}\rangle$. The choice of the PS partial waves
$|\tilde{\phi}_{i}\rangle$ determines the
local contribution of $\hat{\mathcal{T}}_{R}$ to the transformation
operator $\mathcal{T}$, i.e.,
\begin{equation}
|\phi_{i}\rangle=(1+\hat{\mathcal{T}}_{R})|\tilde{\phi}_{i}
 \rangle,\quad\mathrm{within}\ \Omega_{R}.
\end{equation}
For the transformation
$\mathcal{T}$ to be linear, the expansion coefficients have to be
linear functionals of the PS wave functions, i.e., inner products
$\langle\tilde{p}_{i}|\tilde{\Psi}\rangle$ with suitable localized and
fixed projector
functions $\langle\tilde{p}_{i}|$. The transformation $\mathcal{T}$
can now be written as
\begin{equation}\label{transformation}
\mathcal{T}=1+\sum_{i}(|\phi_{i}\rangle-|\tilde{\phi}_{i}\rangle)\langle\tilde{p}_{i}|.
\end{equation}
The final expression for the AE wave function is then
\begin{equation}\label{AEwavefunction}
|\Psi\rangle=|\tilde{\Psi}\rangle+\sum_{i}(|\phi_{i}\rangle-|\tilde{\phi}_{i}\rangle)\langle\tilde{p}_{i}|\tilde{\Psi}\rangle.
\end{equation}
The total energy functional is modified using
Eq.~(\ref{AEwavefunction}). The PS wave functions
$|\tilde{\Psi}\rangle$ are the variational quantities in the PAW
method, which leads to a PS Hamilton operator $\tilde{H}$ acting
on the PS wave functions $|\tilde{\Psi}\rangle$. The soft PS wave
functions obtained can be represented with a modest number of
plane-waves, because there is no norm-conservation requirement for
them and the transformation $\mathcal{T}$ leads to the modified
orthogonality condition
$\langle\tilde{\Psi}_{n}|\mathcal{T}^{\dag}\mathcal{T}|\tilde{\Psi}_{m}\rangle=\delta_{nm}$.
The transformation $\mathcal{T}$
incorporates the effects of the rapid oscillations of the AE wave
functions within the augmentation regions $\Omega_{R}$ into every
expression in the PAW method.

The AE partial waves $|\phi_{i}\rangle$ have to be
orthogonalized to the core states when necessary so that the transformation
$\mathcal{T}$ produces only wave functions orthogonal to the core
electrons. The soft PS partial waves $|\tilde{\phi}_{i}\rangle$
must match the AE partial waves $|\phi_{i}\rangle$ outside
$\Omega_{R}$. The projector functions $\langle\tilde{p}_{i}|$
are chosen so that the completeness
condition $\sum_{i}|\tilde{\phi}_{i}\rangle\langle\tilde{p}_{i}|=1$ is
fulfilled within every $\Omega_{R}$. This enables one to expand
the PS wave function $|\tilde{\Psi}\rangle$
into PS partial waves $|\tilde{\phi}_{i}\rangle$ locally within
every $\Omega_{R}$. The above condition implies that
$\langle\tilde{p}_{i}|\tilde{\phi}_{j}\rangle=\delta_{ij}$.
The construction of the partial waves and projector functions used in this
work is described in detail in Ref.~\cite{Kresse99}. In
practice, the partial wave expansions are truncated. Two reference
energies $\epsilon_{kl}$ are used for each angular momentum quantum
number $l=0,1$ ($l=0,1,2$ for heavy alkali, alkali earth, and d elements).

Further details of the PAW method can be found in
Refs.~\cite{Bloechl94} and \cite{Bloechl03}. The
implementation used in
this work is based on the plane-wave code Vienna
\textit{Ab-initio} Simulation Package~\cite{Kresse96a,Kresse96b}
(\textsc{vasp}) and its implementation of the PAW method~\cite{Kresse99}.

\subsection{PAW valence electron momentum density}

The calculation of the momentum density of the AE wave
functions $|\Psi\rangle$ is straightforward. The PS wave function
part $|\tilde{\Psi}\rangle$ of Eq.~(\ref{AEwavefunction}) is by default
represented on a regular grid in the Fourier space. Note that
$|\tilde{\Psi}\rangle$ extends also to the augmentation regions
$\Omega_{R}$, which  simplifies the calculation in comparison to the
LAPW method. The localized partial waves are functions on a radial grid
multiplied by spherical harmonics $Y_{l}^{m}(\mathbf{\hat{r}})$. The
calculation of  their Fourier transformations is simple,
\begin{eqnarray}
\phi_{i=lmkR}(\mathbf{r}) & = &
\phi_{lk}(|\mathbf{r}-\mathbf{R}|)Y_{l}^{m}(\widehat{\mathbf{r}-\mathbf{R}})\quad\nonumber\\
\stackrel{\mathcal{F}}{\Longrightarrow}\quad\phi_{i=lmkR}(\mathbf{p}) &
= & e^{-i\mathrm{p}\cdot\mathrm{R}}\phi_{lk}(|\mathbf{p}|)Y_{l}^{m}(\mathbf{\hat{p}}).
\end{eqnarray}
The radial part in the Fourier space can be written as
\begin{equation}
\phi_{lk}(p) = \frac{1}{(2\pi)^{3/2}}4\pi(-i)^{l}\int_{0}^{\infty}\mathrm{d}r\,r^{2}\phi_{lk}(r)j_{l}(pr),
\end{equation}
where $j_{l}(pr)$ is a spherical Bessel function. One is now able to
write down the expansion of the AE wavefunction into plane-waves,
\begin{equation}\label{expansion}
\psi_{j\mathbf{k}}(\mathbf{r})=\langle\mathbf{r}|\Psi_{j\mathbf{k}}\rangle=
\frac{1}{\Omega^{1/2}}\sum_{\mathbf{G}}C_{j\mathbf{k}}(\mathbf{G})\exp(i(\mathbf{k}+\mathbf{G})\cdot\mathbf{r}),
\end{equation}
where the expansion coefficient is
\begin{equation}\label{coefficient}
C_{j\mathbf{k}}(\mathbf{G})=\tilde{C}_{j\mathbf{k}}(\mathbf{G})
+\sum_{i}(\phi_{i}\mathbf{(}\mathbf{k}+\mathbf{G})-\tilde{\phi}_{i}(\mathbf{k}+\mathbf{G})\mathbf{)}\langle\tilde{p}_{i}|\tilde{\Psi}_{j\mathbf{k}}\rangle.
\end{equation}
Here $j$ is the band index and $\mathbf{k}$ the Bloch wave vector of
the state, $\Omega$ is the
volume of the supercell, $\mathbf{G}$'s are the reciprocal lattice points
(of the superlattice), and
$\tilde{C}_{j\mathbf{k}}(\mathbf{G})$'s the plane-wave expansion
coefficients of the corresponding PS wave function
$|\tilde{\Psi}_{j\mathbf{k}}\rangle$. The coefficients
$\langle\tilde{p}_{i}|\tilde{\Psi}_{j\mathbf{k}}\rangle$ in
Eq.~(\ref{coefficient}) can be evaluated either in the Fourier space
or in the real space depending on the implementation of the code used in the
computation. Because the plane-waves are eigenfunctions of the momentum,
the momentum density of the state of Eq.~(\ref{expansion}) is
discrete so that the momentum has only values $\mathbf{k}+\mathbf{G}$ with
the probabilities determined by $|C_{j\mathbf{k}}(\mathbf{G})|^{2}$.

In the calculation of electron momentum densities it is not
required to increase the kinetic energy cutoff of the PS wave
functions $|\tilde{\Psi}\rangle$ from typical values of about $250-400$~eV. The
convergence of the AE wave function requires only the
convergence of the PS wave function.
The summation in Eq.~(\ref{AEwavefunction}) converges when the
expansion coefficients $\langle\tilde{p}_{i}|\tilde{\Psi}\rangle$
have converged and the partial waves form a complete set of
functions. Furthermore, it can be shown that the PAW basis set is
complete whenever the plane waves form a complete set, irrespective of
the partial-wave truncation~\cite{Bloechl94}. The truncation, however,
affects the core-valence orthogonality. It also causes errors in
the expectation values and the total energy. In Eq.~(\ref{coefficient}) the
contribution of the PS wave function term
$\tilde{C}_{j\mathbf{k}}(\mathbf{G})$ to
the momentum density extends up to the momentum value
$G_{\mathrm{max}}^{\mathrm{PS}}$ corresponding the kinetic energy
cutoff of the PS wave function. The effect of the partial waves can be
taken into account up to an arbitrary value
$G_{\mathrm{max}}^{\phi-\tilde{\phi}}$.

We have tested the PAW basis set by calculating valence Compton
profiles for several different isolated atoms and compared our results
with the results of an AE code for atoms. We get a perfect match also at
high momenta ($p_{z}>5$~a.u.). The agreement improves systematically with
increasing $G_{\mathrm{max}}^{\phi-\tilde{\phi}}$.

\subsection{Compton profiles}

In the Compton scattering experiment the quantity of interest is the
so-called Compton profile~\cite{Cooper04}
\begin{equation}\label{Compton}
J(p_{z})=\int\int \mathrm{d}p_{x}\,\mathrm{d}p_{y}\,\rho(\mathbf{p}),
\end{equation}
where $\rho(\mathbf{p})$ is the three-dimensional ground-state
momentum density given within the independent-particle model by
\begin{equation}\label{EMD}
\rho(\mathbf{p})=\frac{1}{(2\pi)^{3}}\sum_{j}\bigg |\int \mathrm{d}\mathbf{r}\exp(-i\mathbf{p}\cdot\mathbf{r})\psi_{j}(\mathbf{r})\bigg |^{2},
\end{equation}
where $\psi_{j}(\mathbf{r})$'s are the occupied single-particle
states. Eq.~(\ref{Compton}) is based on the impulse
approximation~\cite{Eisenberger70}. It is assumed to be valid
when the energy transferred in the scattering process is much larger
than the binding energy of the electronic states involved.

In practice we calculate the momentum density of Eq.~(\ref{EMD}) by first
calculating the wavefunctions using a uniform $\mathbf{k}$-point mesh dense
enough to give a sufficient resolution. Then we obtain the momentum
density by summing up the squares of the Fourier coefficients as
\begin{equation}\label{rhopw}
\rho(\mathbf{p})=\frac{1}{\Omega}\sum_{\mathbf{k}}\sum_{j}\sum_{\mathbf{G}}|C_{j\mathbf{k}}(\mathbf{G})|^{2}\delta_{\mathbf{k}+\mathbf{G},\mathbf{p}}.
\end{equation}

The standard way to take into account correlation effects beyond the
independent-particle model in the DFT is the so-called Lam--Platzman (LP)
correlation correction~\cite{Lam74}, which is isotropic within the
LDA. In general, however, comparisons have shown that there are
systematic anisotropic discrepancies between theory and
experiment. As already noted, momentum densities have been calculated
using the QMC methods~\cite{Kralik98,Filippi99} and the $GW$
approximation~\cite{Kubo95,Kubo97,Eguiluz00}. However, in the case of
Si~\cite{Kralik98}, and Li~\cite{Filippi99} (QMC) and
Li~\cite{Schulke99,Eguiluz00} ($GW$) the correlation
correction is not found to differ appreciably from the LP corrected
LDA results~\cite{Schulke04}. Wakoh \textit{et al.}~\cite{Wakoh00}
have shown that the theoretical Compton profiles of Al can be fitted
to the experiment using a phenomenological energy-dependent occupation
number instead of the LP correction. Recently, Barbiellini and
Bansil~\cite{Barbiellini01} have suggested a BCS-like approach in
which the many-body wavefunction is constructed from singlet pair
wavefunctions (geminals) by taking an antisymmetrized geminal product
(AGP).
The scheme is promising in being rather successful in explaining
discrepancies between experiment and previous correction
schemes~\cite{Schulke04}.
Corrections to both the
impulse approximation and to the independent particle model have been
recently proposed within the Dyson orbital formalism~\cite{Kaplan03}.

\subsection{Computational details}

We use in our calculations the LDA based on the
Ceperley--Adler~\cite{Ceperley80} electron gas results as parametrized
by Perdew and Zunger~\cite{Perdew81}, except in the case of the
H$_{2}$O dimer for which a gradient-corrected
functional~\cite{Perdew91} is used. We
model the structures using cubic supercells. The Brillouin zone is
sampled using uniform $\Gamma$-centered
$\mathbf{k}$-point meshes. For the cubic solids we use
the irreducible 1/48th part of the Brillouin zone. We employ the
frozen-core approximation and
calculate only the valence electron momentum density. The core
electron momentum density is isotropic and thus does not affect
anisotropy curves. However, in the case of the bulk Li also the core
electrons are treated self-consistently and taken into account in the
Compton profiles. We do not include the LP correction or any
anisotropic correlation
correction in the calculated Compton profiles. Since the LP correction
is isotropic within the LDA its effect cancels out in the anisotropy
curves. The Compton profiles are calculated by integrating the
three-dimensional electron momentum density directly over planes in the
$\mathbf{p}$-space. The theoretical profiles are convoluted with a
Gaussian function with a full width at half maximum (FWHM)
corresponding to experimental resolution.

\section{Results and discussion}

\subsection{Bulk Si}

We begin by studying the anisotropy of Compton profiles of Si and
compare our results with experiment and a previous computational
study. Delaney \textit{et al.}~\cite{Delaney98} have compared the
results of their PS wave function calculations to experimental Compton
profiles. Further, they have reconstructed AE valence
wave functions from the PS valence wave functions and studied the
effect of the reconstruction on the Compton profiles obtained. Delaney
\textit{et al.}\ used the LDA, norm-conserving
pseudopotentials~\cite{Hamann79} and the
reconstruction scheme by Meyer \textit{et al.}~\cite{Meyer95}. They
noticed that the reconstruction improves the
agreement with experiment substantially. The changes in the momentum
density caused by the reconstruction are not spherically-symmetric and
affect the anisotropy between Compton profiles along different directions
improving the agreement with the experiment. We make the same observation
and get practically identical results with Delaney \textit{et al.}\
when we calculate anisotropies using the AE
and the PS wave functions of the PAW method.
For the bulk Si we use the experimental lattice constant of
10.26~a.u. The $\mathbf{k}$-point mesh used is a $\Gamma$-centered
$10\times 10\times 10$ mesh. The sampling corresponds
to the momentum resolution of 0.061~a.u. The kinetic energy cutoff for
the PS wave functions is  307~eV. The three-dimensional momentum
density used in the calculation of the Compton profiles
[Eq.~(\ref{rhopw})] covers momenta up to $p_{\mathrm{max}}=5.0$~a.u.

\begin{figure}
\begin{center}
\includegraphics[width=.8\textwidth]{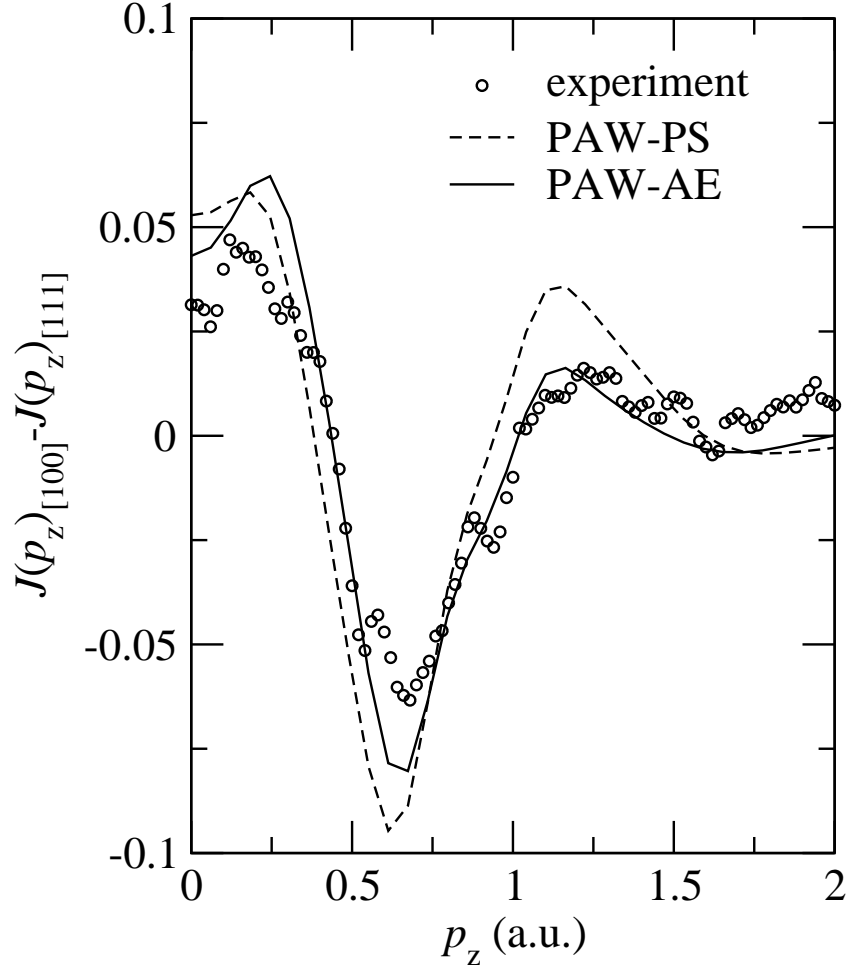}
\caption{Bulk Si. The anisotropy between the Compton
  profiles along the [100] and [111] directions. The theoretical
  curves are convoluted with a Gaussian function with a FWHM of 0.13
  a.u.\ corresponding to the experimental resolution.  The
  experimental data is from Ref.~\cite{Kubo97b}.}\label{anisotropy_Si}
\end{center}
\end{figure}

Figure~\ref{anisotropy_Si}
shows our results for the [100]--[111] anisotropy compared with the
experimental result from Ref.~\cite{Kubo97b}. The transformation
$\mathcal{T}$ from the PS wave functions to the AE wave functions
reduces the overestimation of the amplitudes of the peaks in the
anisotropy. The [100]--[110] anisotropy is also improved but
relatively less affected by the reconstruction.


\subsection{Bulk Li}

Bulk Li has been the subject of many high-resolution Compton
scattering studies. Its Fermi surface has been studied by looking at
the first and second derivatives of directional Compton
profiles~\cite{Sakurai95} and by reconstructing the full
three-dimensional momentum density~\cite{Schulke96,Tanaka01}. We
compare our results for Li with experimental high-resolution
Compton profiles and the theoretical results from
Ref.~\cite{Sakurai95}. The theoretical results in
Ref.~\cite{Sakurai95} have been calculated with the KKR band
structure scheme and the LDA. The momentum density of Li has also been
calculated using the LMTO method~\cite{Manuel93,Dugdale98}, the FLAPW
method~\cite{Kubo95,Kubo97,Baruah99}, the $GW$
approximation~\cite{Kubo95,Kubo97,Eguiluz00}, and quantum Monte Carlo
methods~\cite{Filippi99}. Theoretical calculations based on the
LDA have failed to describe the momentum density of Li also when the
Lam--Platzman correlation correction has been taken into
account~\cite{Sakurai95}. The discrepancies have been attributed to
electronic correlation beyond the LDA~\cite{Sakurai95,Schulke96} or to
the thermal disorder~\cite{Dugdale98}. Our calculation does not
properly take into account these effects. Therefore, instead of
experimental data we have to use other computational results based on
the LDA as benchmarks.

We calculate our results for the bulk Li using the experimental
lattice constant of 6.60~a.u. The
$\mathbf{k}$-point mesh used is a $\Gamma$-centered $30\times 30\times
30$ mesh. The sampling corresponds to the momentum resolution of
0.032~a.u. The kinetic energy cutoff of the PS wave functions is
271~eV. Momenta up to $p_{\mathrm{max}}=10.0$~a.u.\ are taken into account when
calculating the Compton profiles.

\begin{figure}
\begin{center}
\includegraphics[width=.6\textwidth]{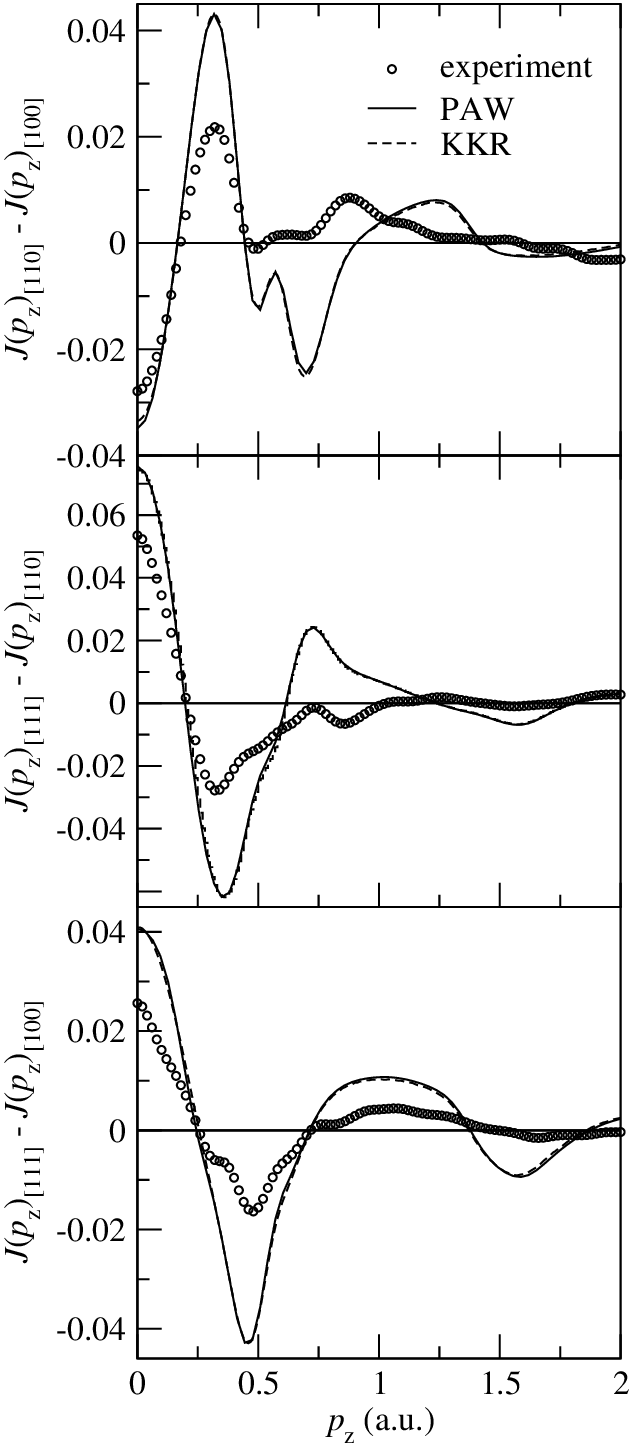}
\caption{Bulk Li. Anisotropies between Compton
  profiles along different directions. The theoretical
  curves are convoluted with a Gaussian function with a FWHM of 0.12
  a.u.\ corresponding to the experimental resolution. The experimental
  data and the KKR result are from
  Ref.~\cite{Sakurai95}.}\label{anisotropies_Li}
\end{center}
\end{figure}

Figure~\ref{anisotropies_Li} shows in addition to our results experimental
anisotropies~\cite{Sakurai95} and those calculated with the
KKR~\cite{Sakurai95}. The PAW and the KKR results are in perfect
agreement although the muffin-tin approximation has been used in the
KKR calculation. A disagreement is found between our results and the
FLAPW results by Baruah \textit{et al.}~\cite{Baruah99}\ although in
principle the only major difference between the calculations is the
choice of the basis set. Actually, Baruah \textit{et al.}\ used
gradient corrections to the exchange-correlation potential, but
concluded that their effect on the electron momentum density is not
significant. However, the FLAPW results by Kubo~\cite{Kubo95} are in
good agreement with the KKR method, and thus also agree with our
calculation.


\subsection{Bulk Al}

Ohata \textit{et al.}~\cite{Ohata00} have measured directional
high-resolu\-tion Compton profiles of Al and compared them with
theoretical results calculated with the KKR band structure scheme within
the LDA. For Al the agreement between theory and
experiment was found to be better than in the case of Li. Correlation
effects are less important as the electron density is higher. Al is
also harder than Li and other alkali metals, which reduces the effect
of thermal disorder on Compton profiles. However, because the
anisotropy of Compton profiles of Al is very
small, and the anisotropy curves and derivatives of Compton profiles
contain detailed features, Al is the most challenging test for the PAW
method presented in this paper.

For the bulk Al we use the experimental
lattice constant of 7.65~a.u. The $\mathbf{k}$-point mesh used is a
$\Gamma$-centered $60\times 60 \times 60$ mesh. The sampling
corresponds to the momentum resolution of 0.014~a.u. The kinetic
energy cutoff for the PS wave functions is 296~eV. Momenta up to
$p_{\mathrm{max}}=5.0$~a.u.\ are taken into account when calculating
the Compton profiles.

\begin{figure}
\begin{center}
\includegraphics[width=.9\textwidth]{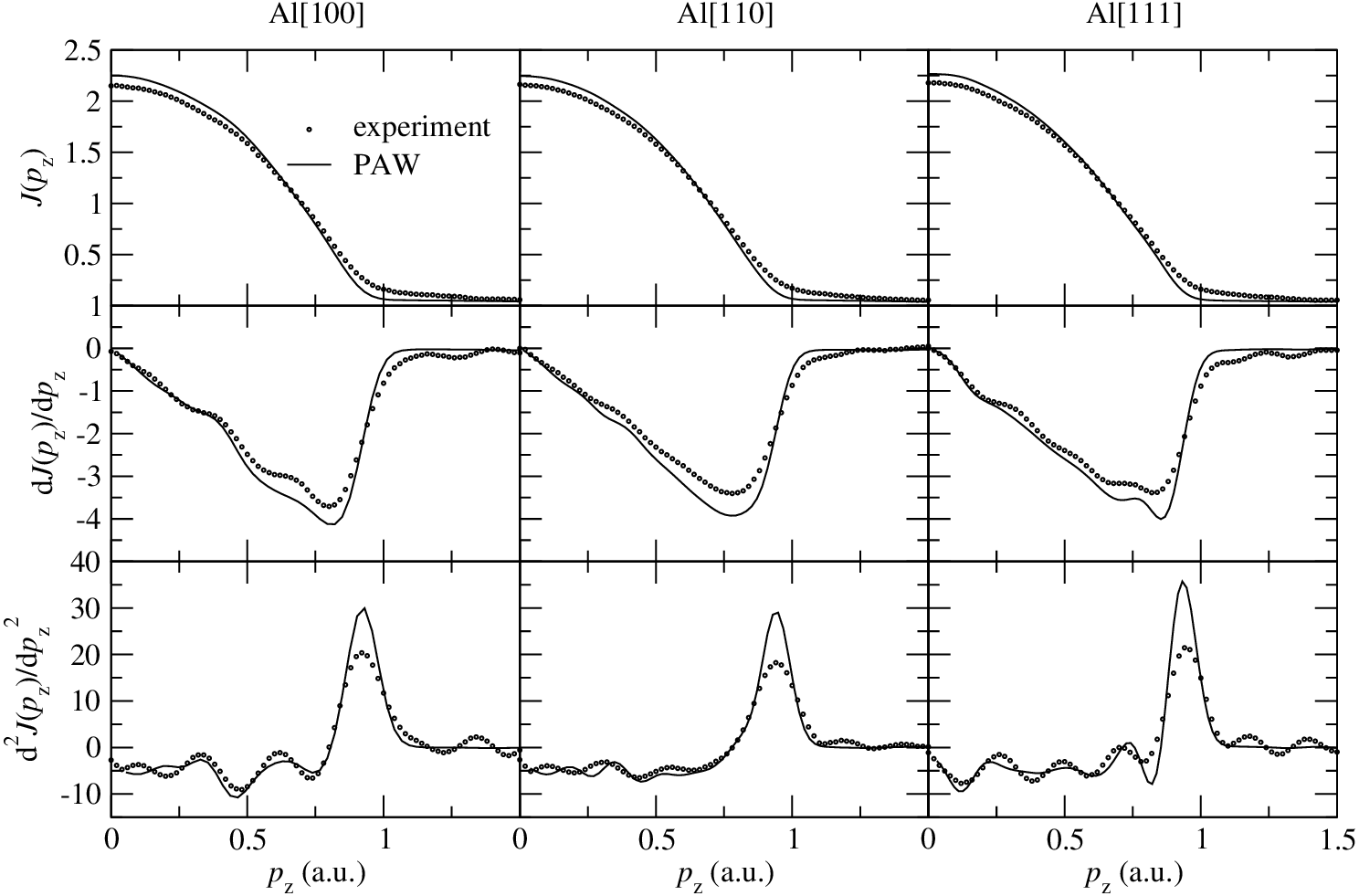}
\caption{Valence Compton profiles of Al
  along different directions (the first row). Also
  their first and second derivatives obtained by numerical
  differentiation are shown (the second and third rows). The theoretical
  Compton profiles are convoluted with a FWHM of 0.12 a.u.\
  corresponding to the experimental resolution. The experimental valence
  Compton profiles (from Ref.~\cite{Ohata00}) have been obtained
  by subtracting the KKR core profile from the measured
  ones.}\label{derivatives_Al}
\end{center}
\end{figure}
\begin{figure}
\begin{center}
\includegraphics[width=.6\textwidth]{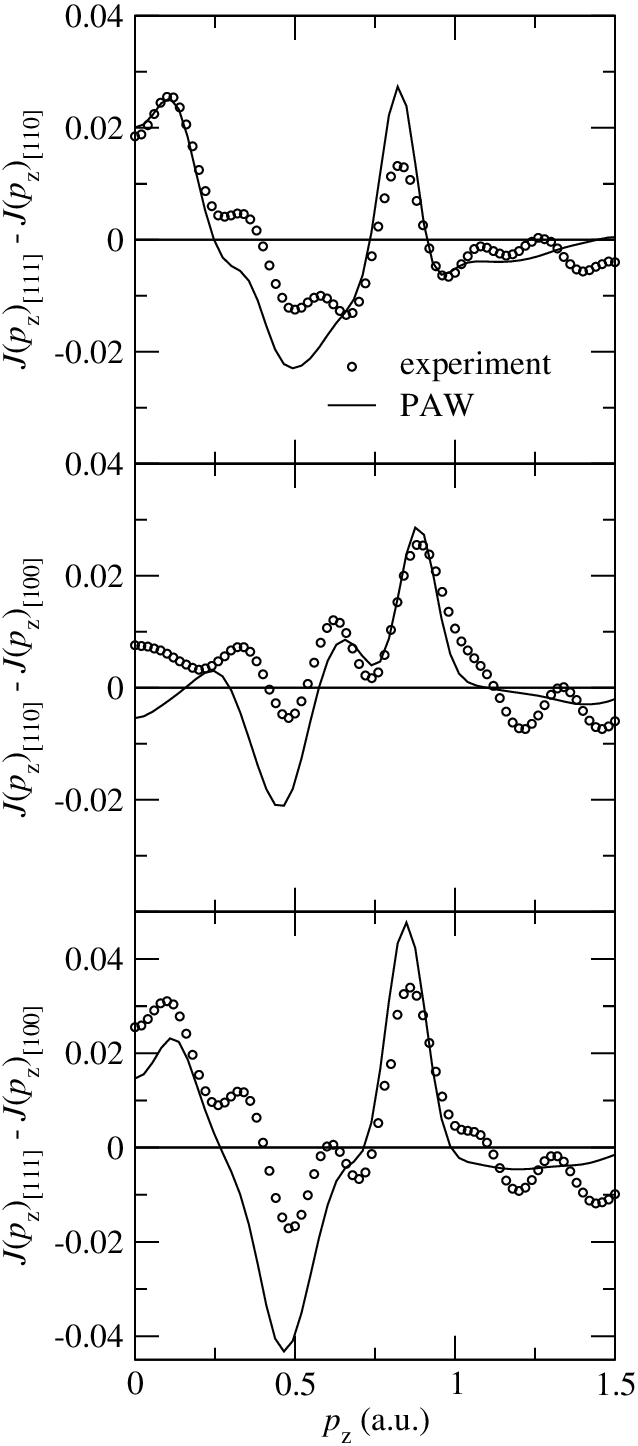}
\caption{Bulk Al. Anisotropies between Compton
  profiles along different directions. The theoretical
  curves are convoluted with a Gaussian function with a FWHM of 0.12
  a.u.\ corresponding to the experimental resolution. The experimental
  data is from Ref.~\cite{Ohata00}.}\label{anisotropies_Al}
\end{center}
\end{figure}

Figure~\ref{derivatives_Al} shows the calculated valence Compton
profiles and their first and second derivatives compared with the
experimental results of
Ref.~\cite{Ohata00}. As
can be expected, for the LDA, the theoretical Compton profiles are
higher at low momenta~\cite{Sakurai95,Delaney98,Ohata00}. The
Lam--Platzman correlation correction would shift weight
from lower to higher momenta improving the agreement with the
experiment. The first and second derivatives of the
Compton profiles calculated with the
PAW method agree better with experiment than the KKR
results by Ohata \textit{et al.}\ (see the data in
Ref.~\cite{Ohata00}). Anisotropies of the Compton profiles are studied in
Fig.~\ref{anisotropies_Al}. Agreement with the KKR scheme is not as
good as in the case of Li. Compared with the experiment the agreement
is about as good for both theoretical methods. The PAW method does
not reproduce all the features of the
experimental anisotropies as clearly as the KKR scheme. However, it
predicts the positions of the peaks in the anisotropies
better, especially in the [110]--[100] and
[111]--[100] differences.


\subsection{Water dimer (H$_{2}$O)$_{2}$}

Finally, we calculate the Compton profile of the gas-phase water dimer
using the generalized-gradient approximation (GGA) by Perdew and
Wang~\cite{Perdew91}, and the experimental geometry by Dyke \textit{et
al.}~\cite{Dyke77}\ with the O--O separation of 2.98~\AA. The water
dimer can be considered as a prototype system to study how the
constitutive interactions of the hydrogen bond (e.g.\ exchange
repulsion, charge transfer and polarization) are reflected in the
Compton profile~\cite{Ghanty00}. The gas-phase dimer
represents a moderately hydrogen-bonded system, where the main
oscillatory signal in the Compton profile is expected to stem from the
exchange interaction of the overlapping molecular wave
functions~\cite{Ghanty00}. Since the binding compared to the metallic
and covalent systems is much weaker, the wave functions of the two
water molecules conserve to a large extent their unperturbed
form. Hydrogen bonds provide thus one of the limiting cases to study
the applicability of the PAW calculations to describe chemical
bonding. The PAW result is calculated using a
periodic supercell with a side length of 35.0~a.u.\ in order to
minimize the interaction between the periodic images of the dimer. The
\textbf{k}-point sampling (a $\Gamma$-centered $3\times 3\times 3$ mesh)
used when calculating the momentum density corresponds to the momentum
resolution of 0.060~a.u. In the
calculation of the effective potential, however, only the $\Gamma$
point is used. The kinetic energy cutoff of the PS wave functions is
500~eV. The momentum density is taken into account up to the
momentum value of $p_{\mathrm{max}}=13.7$~a.u. when calculating the
Compton profile. We calculate the Compton profile of the H$_{2}$O
dimer also by employing localized basis functions. The same GGA
functionals are used. For oxygen a triple-zeta valence plus
polarization type basis set is used, and for hydrogen a primitive set
augmented by one p-function in a [3s, 1p] contraction (for more
details, see Ref.~\cite{Hakala04}). No experimental data is available
for the directional anisotropies of the dimer.

Figure 5 shows the anisotropy between the profile along the O--O
direction (direction of the H-bond) and that in the direction
orthogonal to the mirror plane of the H$_{2}$O dimer. The solid line
denotes the AE PAW result, the dashed line the result
calculated with localized basis functions, and the dotted line the
result calculated using the PS wave functions of the PAW method.
Agreement between the two AE calculations is very good,
especially at high momenta. The positions of the peaks are improved
discernibly when the PS wave functions are replaced by the
AE wave functions. At momenta $p_{z}>2.5$~a.u. the PS wave
functions fail to describe the positions of the peaks properly. At lower
momenta the deficiencies in the basis sets and the residual
interaction between the periodic images in the PAW method explain the
slight differences in the amplitudes between the AE calculations.


\section{Summary and conclusions}

We have presented valence electron Compton profiles calculated within
the DFT using the projector augmented-wave (PAW) method. The accuracy
of the method has been demonstrated to be competitive with other
band-structure and basis set schemes. The PAW method is ideal for
the calculation of valence electron momentum densities of covalent
systems. Also for systems with much weaker orbital rearrangements,
such as hydrogen bonds, the PAW basis set yields reliable results. It
becomes highly beneficial to use a unified approach with a
well-controlled basis set and full-potential treatment for the wide
variety of nonmetallic systems. The effect of the frozen-core
approximation is negligible in most systems. The quality of the PAW
calculations can be systematically improved by increasing the energy
cutoff parameter. For metallic systems (Li, Al) in the LDA, despite
the coarser \textbf{k}-mesh, PAW yields essentially as good results as
the KKR method. Nevertheless, for metals both methods suffer from
discrepancies as compared with experiment, attributed to electron
correlations beyond the LDA or to thermal disorder. Our work
suggests that the PAW method which is becoming ever more popular in
electronic structure calculations suits well for describing a wide
variety of ordered and disordered covalently or hydrogen bonded systems.

\begin{ack}
We thank Dr.\ Y.\ Sakurai for sending the experimental data on Si, Li
and Al, and Prof.\ A.\ Bansil for providing the KKR Compton profiles
of Li. We also thank Prof.\ S.\ Manninen, Prof.\ K.\ H\"am\"al\"ainen, and
Academy Prof.\ R.\ Nieminen for discussions. We acknowledge the
generous computer resources from the Center of Scientific Computing,
Espoo, Finland. This work has been supported by the Academy of Finland
through its Centers of Excellence program (2000-2005) and contract No.\
205967/201291 (MH).
\end{ack}




\bibliographystyle{elsart-num}
\bibliography{pawpaper}
\clearpage

\begin{figure}
\begin{center}
\includegraphics[width=.8\textwidth]{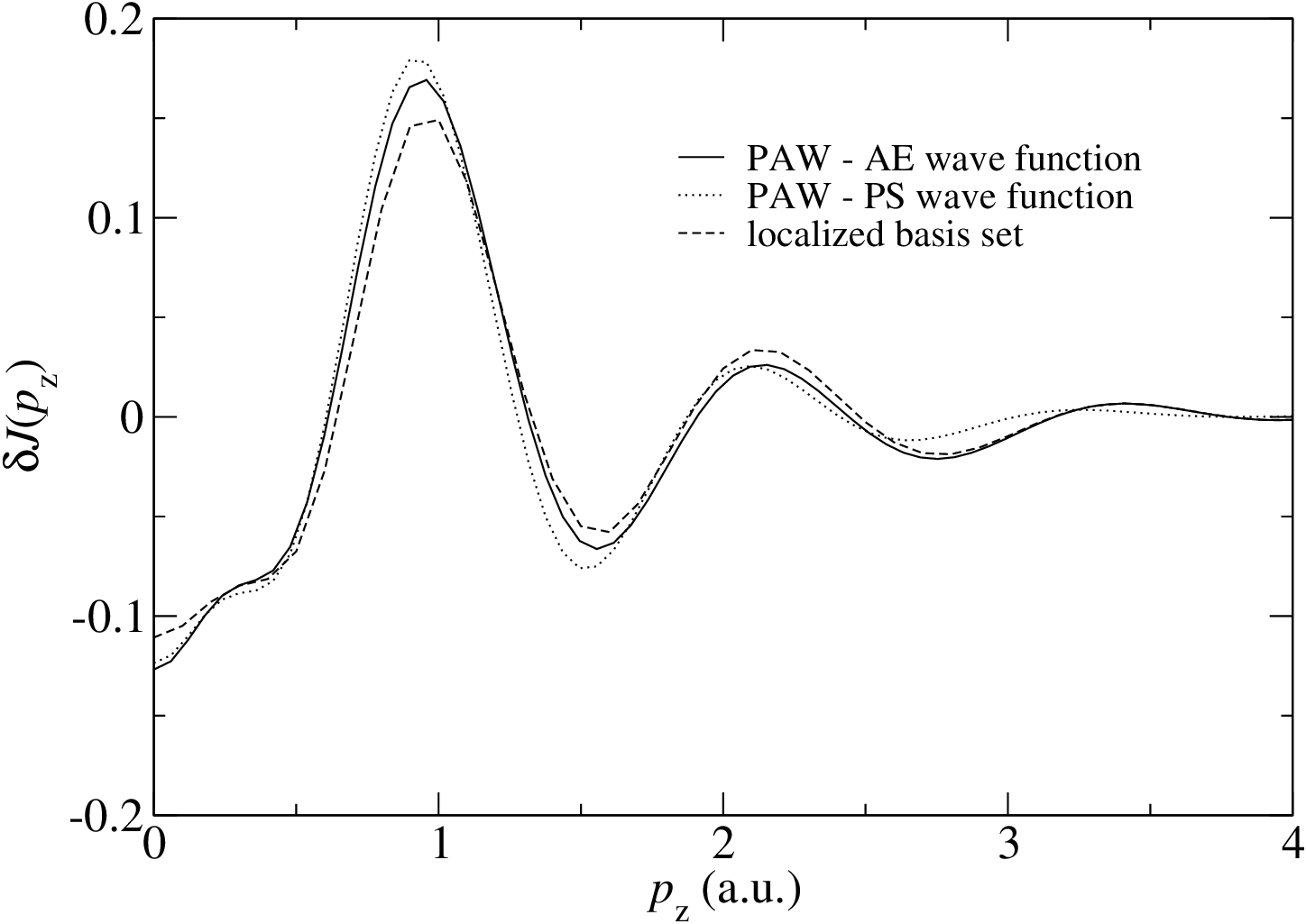}
\caption{Compton profile anisotropy for the gas-phase water
  dimer.}\label{dimer}
\end{center}
\end{figure}
\end{document}